\documentclass{aa}  

\usepackage{graphicx}
\usepackage{txfonts}
\usepackage{times}
\usepackage{epsfig}
\usepackage{lscape}
\usepackage{amssymb}
\usepackage{latexsym}
\usepackage{natbib}
\def\kms{$\rm km\;s^{-1}$}

\def\hb{H$\beta$}

\def\oiiipg{[O~{\small III}]$\,\lambda\lambda4959,5007\AA$}
\def\oiiip{[O~{\small III}]$\,\lambda4959$}
\def\oiiig{[O~{\small III}]$\,\lambda5007\AA$}

\def\mfe{$\langle$Fe$\rangle$}

\def\mgfe{[MgFe]'}

\def\mgb{Mg~{\it b}}
\def\fei{Fe{\small 5270}}
\def\feii{Fe{\small 5335}}
\def\Msun{$M_\odot$}

\begin{document}

   \title{The properties and the formation mechanism of the stellar counter-rotating components in NGC 4191}

   \subtitle{}

   \author{L. Coccato \inst{1}
          \and M. Fabricius \inst{2,3,4}
          \and L. Morelli \inst{5,6} 
          \and E. M. Corsini \inst{5,6}
          \and A. Pizzella \inst{5,6} 
          \and P. Erwin \inst{2,3}
          \and E. Dalla Bont\`a \inst{5,6} 
          \and R. Saglia \inst{2,3}
          \and R. Bender \inst{2,3}
          \and M. Williams \inst{2,7}
          }
  \offprints{L. Coccato, e-mail: lcoccato@eso.org}

  \institute{European Southern Observatory, Karl-Schwarzschild-Stra$\beta$e 2, 
            D-85748 Garching bei M\"unchen, Germany.  
       \and Max Planck Institute f\"ur Extraterrestrial Physics, Giessenbachstra$\beta$e, 
            D-85748 Garching, Germany.
       \and University Observatory Munich, Scheinerstra$\beta$e 1, 81679 Munich, Germany.
       \and Subaru Telescope, 650 North Aohoku Place, Hilo, HI 96720.
       \and Dipartimento di Fisica e Astronomia ``G. Galilei'',
            Universit\`a di Padova, vicolo dell' Osservatorio 3, I-35122
            Padova, Italy. 
       \and INAF-Osservatorio Astronomico di Padova, vicolo dell' Osservatorio 5, 
            I-35122 Padova, Italy.
       \and Department of Astronomy, Columbia University, New York, NY 10027, USA. 
    }

   \date{Received April 2015; accepted July 2015}

 
  \abstract 
  {Formation of stellar counter-rotating components in galaxies.}
  {In this paper we disentangle two counter-rotating
    stellar components in NGC 4191 and characterize their physical
    properties such as kinematics, size, morphology, age, metallicity,
    and abundance ratio. NGC 4191 was selected as candidate to host
    stellar counter-rotating components on the basis of two symmetric
    peaks in its velocity dispersion two-dimensional map and its
    irregular velocity field.}
  {We obtained integral field spectroscopic observations with VIRUS-W
    and performed a spectroscopic decomposition technique to separate
    the contribution of two stellar components to the observed galaxy
    spectrum across the field of view.  We also performed a
    photometric decomposition, modeling the galaxy with a S\'ersic bulge
    and two exponential disks of different scale length, with the aim
    of associating these structural components with the kinematic
    components.  We then measured the equivalent width of the
    absorption line indices on the best fit models that represent the
    kinematic components and compared our measurements to the
    predictions of stellar population models that account also for
    variable abundance ratio of $\alpha$ elements.}
  {We have evidence that the line-of-sight velocity distributions
    (LOSVDs) are bimodal and asymmetric, consistent with the presence
    of two distinct kinematic components. The combined information of
    the intensity of the peaks of the LOSVDs and the photometric
    decomposition allows us to associate the S\'ersic bulge and the
    outer disk with the main kinematic component, and to associate the
    inner disk with the secondary kinematic component. We find that
    the two kinematic stellar components counter-rotate with respect
    to each other. The main component is the most luminous and
    massive, and it rotates slower than the secondary component, which
    rotates along the same direction as the ionized gas. The study of
    the stellar populations reveals that the two kinematic components
    have the same solar metallicity and sub-solar abundance ratio,
    without the presence of significant radial gradients. On the other
    hand, their ages show strong negative gradients and the
      possible indication that the secondary component is the
      youngest. We interpret our results in light of recent
    cosmological simulations and suggest gas accretion along two
    filaments as the formation mechanism of the stellar
    counter-rotating components in NGC 4191.}
  {}


   \maketitle
%

\section{Introduction}

The phenomenon of counter-rotation, i.e. the presence of multiple
kinematic components that are rotating in opposite directions, has
been detected in a number of galaxies of all morphological types (see
\citealt{Corsini14} for a review). Counter-rotating galaxies are
classed according to the nature of their counter-rotating components,
i.e., gas vs. stars (e.g., NGC 4546, \citealt{Galletta+87}), stars vs.
stars (e.g., NGC 4550, \citealt{Rubin+92, Rix+92}), and gas vs. gas
(e.g., NGC 7332, \citealt{Fisher+94}). For the cases of stellar
counter-rotation, a further classification can be done by looking at
the sizes of the decoupled structures. There are galaxies with two
counter-rotating stellar disks of similar sizes (e.g., NGC 4550) and
galaxies where the counter-rotation is visible only in the innermost
regions (e.g., NGC 448, \citealt{Krajnovic+11}).

Thanks to the advent of integral field spectroscopic surveys like
SAURON \citep{DeZeeuw+02}, ATLAS 3D \citep{Cappellari+11},
CALIFA \citep{Sanchez+12}, SAMI \citep{Bryant+14}, and MANGA
\citep{Bundy+15}, the census of counter-rotating galaxies has
increased.  Indeed, these surveys allowed to identify candidate
galaxies to host counter-rotating stellar disks by looking for the
kinematic signature given by two off-center and symmetric peaks in the
stellar velocity dispersion in combination with zero velocity rotation
measured along the galaxy major axis. These kinematic features are
observed in the radial range where the two counter-rotating components
have roughly the same luminosity and their LOSVDs are unresolved
\citep{Rix+92, Bertola+92, Vergani+07}. Recently, \citet{Krajnovic+11}
have found 11 galaxies (including NGC 4191) with a double-peaked
velocity dispersion in the volume-limited sample of 260 nearby
early-type galaxies gathered by the ATLAS-3D project.

The current paradigm that explains stellar counter-rotation is a
retrograde acquisition of external gas and subsequent star formation
\citep{Thakar+96, Pizzella+04, Algorry+14}. Alternative scenarios such
as the assembly of the counter-rotating stellar component from mergers
\citep{Puerari+01, Crocker+09} or internal instabilities
\citep{Evans+94} have been proposed, too. The relatively small number of
studied cases favors external origin, but it does not yet allow us to
distinguish between gas accretion or merging \citep{Coccato+13,
  Pizzella+14}.

In order to characterize the physical properties of the two stellar
counter-rotating components and therefore to constrain their formation
mechanism, we need to disentangle their contribution to the observed
galaxy spectrum. To this aim, we introduced a spectral decomposition
technique that allows one to measure the kinematics and properties of the
stellar populations of the decoupled components \citep{Coccato+11}.

Other parametric \citep{Johnston+13} and non parametric
\citep{Katkov+11} techniques have also been proposed and developed by
other teams.  All these studies show that the stellar component that
is rotating along the same direction as the ionized gas is 
younger, less massive, and has different metallicity and abundance
ratio with respect to the main galaxy component.  These results are
consistent with the gas accretion scenario followed by star
formation. We also found that, at least for one case, the small
counter-rotation observed in the inner regions is the ``tip of the
iceberg'' of a much larger structure, which is less luminous than the
main stellar component and whose real extent can be revealed only by
a spectral decomposition (NGC 3593, \citealt{Coccato+13}).

In this work, we investigate the S0 galaxy NGC 4191, which is an
isolated system located at a distance of 42.4 Mpc \citep{Theureau+07}.
In contrast to the galaxies we studied in our previous works
\citep{Coccato+11,Coccato+13,Pizzella+14}, NGC 4191 was not already
known to host two large-scale counter-rotating stellar disks.  We
selected NGC 4191 because its stellar velocity field reveals an
irregular structure with little rotation, and its stellar velocity
dispersion field shows two symmetric peaks located on opposite sides
at $\sim 10"$ from the galaxy center along the photometric major axis
\citep{Krajnovic+11}. The latter feature is one of the key
observables proposed for detecting candidate galaxies with stellar
counter-rotation.  Contrarily to its complex kinematic structure, the
photometric profile of NGC 4191 does not reveal prominent
sub-structures. The light distribution of NGC 4191 has been modeled
with a single S\'ersic law with index $n=2.4$
\citep{Krajnovic+13}.

The aim of the paper is to study the structure of NGC 4191 and to
investigate the presence of multiple kinematic stellar components.  In
particular, we want to test whether NGC 4191 does host
counter-rotating stellar components and, if present, we want to
characterize their kinematics and stellar population properties. Also,
we want to investigate whether the kinematic components can be
associated with any photometric sub-structure, which might have been
overlooked in previous studies. Moreover, if two counter-rotating
components are indeed detected, we will have strengthened the case
that the presence of a double-peaked signature in the stellar velocity
dispersion field combined with the absence of regular rotation can be
used as selection criteria for counter-rotating galaxies.

This paper is structured as follows; Section
\ref{sec:observations_reduction} describes the spectroscopic
observation and the data reduction; Section \ref{sec:analysis}
presents our analysis and our results. Finally, we summarize our
conclusions in Section \ref{sec:summary}.


\section{Observations and data reduction.}
\label{sec:observations_reduction}

\subsection{Observations}
\label{sec:observations}

The observations of NGC\,4191 were carried on 1-3 and 6-7 April 2013
using the VIRUS-W Integral Field Unit (IFU) Spectrograph
\citep{Fabricius2012b} at the 2.7-m Harlan J. Smith Telescope of the
McDonald Observatory (Texas, US).  The IFU consists of 267 fibres
arranged in a rectangular grid, which covers a field of view $105\arcsec
\times 55\arcsec$ with a fill factor of 1/3. Each fibre has a diameter
of $3 \farcs 2 $ on the sky.  We used the lower resolution mode of the
instrument which covers the 4340\AA\ -- 6040\AA\ wavelength range 
with a spectral resolution of 1.57 \AA\ FWHM ($\sigma_{\rm instr} \sim
$ 38 km\,s$^{-1}$) \footnote{The spectral resolution of VIRUS-W below
  4800 \AA\ deteriorates up to $\sigma_{\rm instr} \sim 70$ \kms. We
  tested that restricting the wavelength range to 4800\AA\ -- 6040\AA\
  has negligible effects on our results.}.

The observations were dithered to fill the entire field of view. We
took 6 exposures of 1800 sec in each of the three dither positions
and bracketed and interleaved them with 300 sec sky nods. Each of the
science exposures was split in two for cosmic ray rejection. The total
on-target exposure time per fibre is 3\,h. We recorded bias frames and
Hg+Ne arc lamp exposures for wavelength calibration on the evenings
and mornings before and after the observations. We also recorded dome
flat exposures to trace the fibre positions on the detector and to
compensate for fibre-to-fibre throughput variation. A set of
spectroscopic standard stars were also observed with the same
instrumental set-up.

\begin{figure}
\begin{center}
\psfig{file=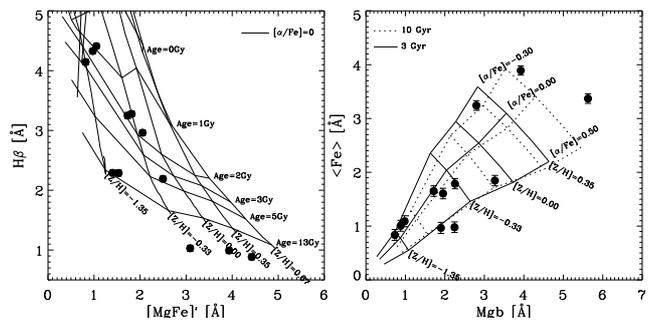,width=4.1cm,clip=,angle=90}
\caption{Location of the observed stars (black dots) in the \mgfe,
  \hb\ (left panel) and the \mgb, \mfe\ (right panel) parameter
  space. The line grids give the predictions from simple stellar
  population models by \citet{Thomas+03}.}
\label{fig:virus_ssp}
\end{center}
\end{figure}

\subsection{Data reduction}
\label{sec:reduction}

The data reduction follows the same methodology as described in
\citet{Fabricius+14}. In short, after subtraction of the master bias the
fiber positions are traced by searching along the peaks of the spectra in the
dome flat frames. Then the arc peak positions are identified in the master arc
frames by searching along those traces. The fiber/wavelength to x/y pixel
position mapping is then modeled and fit --- using a standard least squares
minimization routine --- with a two-dimensional 7-th degree Chebyshev
polynomial. We use 28 Hg and Ne spectral lines for the fit.

After the wavelength calibration we average the two cosmic ray
split science exposures as well as the bracketing sky frames while
rejecting spurious events using the {\tt PyCosmic} routine
\citep{Husemann2012}. We extract science, sky and dome-flat spectra by
stepping along the previously determined trace positions.  For each
step, the counts in the CCD pixels are assigned to spectral pixels
according to the areal overlap of the CCD pixel with the 7 pixel wide
and 0.52\,\AA\ long extraction aperture. We correct for
fiber-to-fiber throughput variations using the extracted dome-flat
spectra and remove the sky background by scaling the sky spectra by
the relative exposure time.

Then, we combine all flat-fielded and background-removed science spectra into
one datacube. We impose a regular grid with $1\farcs 6 \times 1\farcs 6$ large pixels and
calculate the areal overlap between all circular fiber apertures and the
pixels. The flux in each fiber is then distributed to pixels according to the
overlap.

Finally, we spatially rebin the final datacube to increase the
observed signal-to-noise ratio ($S/N$) using the implementation for
Voronoi Tesselation by \citet{Cappellari+03}.

\section{Analysis: kinematics, photometry, and stellar populations}
\label{sec:analysis}
In this Section, we measure the stellar and ionized gas kinematics of
NGC 4191 and its structural components. The stellar library we use is
presented in Section \ref{sec:library}. We first analyze the spectra
with a non-parametric fit to recover the shape of the line-of-sight
velocity distribution (LOSVD) in each spatial bin (Section
\ref{sec:one_comp}). Motivated by the bimodality of the recovered
LOSVDs, we apply a spectral decomposition technique to investigate and
characterize the presence of two kinematically distinct stellar
components (Section \ref{sec:two_comp}). In Sect. \ref{sec:two_comp},
we also study the structural components of NGC 4191 and investigate
the connection between the photometric and kinematics components. We
then measure the properties of stellar populations of the two
kinematic components: line strength indices, age, metallicity, and
abundance ratios of $\alpha$ elements (Section \ref{sec:ssp}). The
distribution of the ionized gas is discussed in Section
\ref{sec:ionized_gas}.

\subsection{The stellar library}
\label{sec:library}

The stellar library used in the spectroscopic fit consists of 12 stars
observed with Virus-W using the same observational set-up of the galaxy
observations. This ensures that the stellar templates had the same
line spread function of the galaxy, minimizing systematic effects due
to mismatch between the resolution of the stars and the galaxy
spectra. This is particularly important for systems with low velocity
dispersion, where the asymmetries in the LOSVDs are comparable with
asymmetries in the line spread function.

The selected stars covered a large range of the \hb, \mgb, \mgfe, and
\mfe\ parameters space, as shown in Figure \ref{fig:virus_ssp}.

\begin{figure}
\begin{center}
\psfig{file=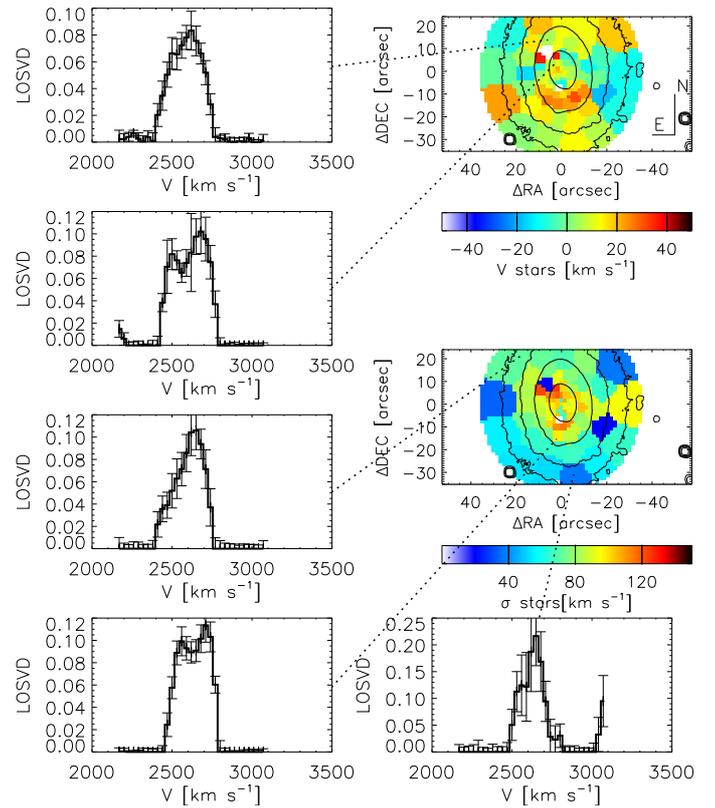,width=9cm,clip=,bb=38 358 426 820}
\caption{Best fit kinematics for the single stellar component
  model. The two-dimensional maps show the velocity and velocity
  dispersion fields. The plots on the side show some example of
  reconstructed non-parametric LOSVDs in the location of the field of
  view indicated by the dotted lines. The LOSVDs of central regions
  clearly show a bimodal distribution; the LOSVDs in the outer regions
  show high asymmetry, which can be due to the presence of an
  unresolved secondary kinematic component.}
\label{fig:one_comp}
\end{center}
\end{figure}

\subsection{Shape of the line-of-sight velocity distributions}
\label{sec:one_comp}

We fitted the spectra with a non-parametric approach in order to
recover the shape of the LOSVDs and investigate the presence of
multiple kinematic components. The fitting procedure is described in
\citet{Fabricius+14}, and it extends the maximum penalized likelihood
method by \citet{Gebhardt2000b} to treat nebular emissions. After
fitting and removing the continuum with a 3-rd degree polynomial, we
fitted the wavelength range from 4905\,\AA - 5325\,\AA. We modeled the
LOSVD with 31 independent velocity channels and linearly combined the
set of 12 stellar spectra from a library (Sect. \ref{sec:library}). We
successively removed stellar spectra that received small weights.  The
final fit was carried out with six templates. Due to the high $S/N$ of
the data after binning (typically 90 per \AA) we did not employ any
penalization. For a number of bins the derived LOSVDs show a clear
bimodal structure expressed as a double peak.

These bifurcated LOSVDs are more evident in the inner regions of the
galaxies ($r<10''$) and are spatially consistent with the regions
where the sigma-peaks were detected. For the
regions outside $10''$ the LOSVDs are strongly asymmetric, consistent
with the interpretation that the galaxy contains two counter-rotating
stellar components.

Figure \ref{fig:one_comp} shows the shape of the LOSVDs in some
spatial bins. In order to measure the mean kinematics along the
line-of-sight, we re-fitted the spectra parametrizing the LOSVD in
each bin with Gaussian function plus high-order Gauss-Hermite
moments. The fit was done using a modified version of the Penalized
Pixel Fitting code of \citet{Cappellari+04} that included simultaneous
fits to the ionized gas emission lines. Fig. \ref{fig:one_comp} shows
the velocity and velocity dispersion fields obtained with the
parametric fit.  The two peaks in the velocity dispersion profile
discovered by \citet{Krajnovic+11} are visible, slightly
  offsetted by $\approx 1''$ towards South East with respect to the
  galaxy photometric major axis. The velocity field is complex and it
is not consistent with a single rotating component, even if high order
moments are taken into account in the LOSVD parametrization, providing
further support to the presence of kinematically decoupled components.

\subsection{Kinematics and photometry of the counter-rotating stellar components}
\label{sec:two_comp}

Driven by the results of Section \ref{sec:one_comp}, we apply a
spectral decomposition technique to investigate and characterize the
presence of two kinematically distinct stellar components. 

The spectral decomposition, described in Section
\ref{sec:decomposition}, returns for each spatial bin the kinematics,
the spectra, and the normalized flux contribution of the two stellar
components. We also fit the intensities of the emission lines, and
their mean velocity and velocity dispersion. We used the same spectral
library for both the components, and let the code to select the most
appropriate stellar template for each of them.

In Section \ref{sec:photometry} we associate these two kinematic
components with the galaxy structural components derived from a
photometric decomposition.

\subsubsection{Spectral decomposition technique}
\label{sec:decomposition}

The parametric fit of the stellar kinematics and the separation of two
kinematic stellar components is done using the spectral decomposition
technique developed in \citet{Coccato+11}, implemented as modification to the
Penalized Pixel Fitting code \citep{Cappellari+04}.

The technique builds two optimal templates by linear combination of
stellar spectra from an input library (see Section
\ref{sec:library}). The two templates are convolved with two
independent LOSVDs, which are parametrized by two independent Gaussian
functions. The convolved templates are then multiplied by Legendre
polynomials to account for the shape of the continuum. Our
implementation includes Gaussian functions that are added to fit the
ionized gas emission lines \hb\ and \oiiipg. The best fit model is
recovered using the Levenberg-Mardquart $\chi^2$ minimization. The
procedure allows the flux of the two components either to be free
parameters in the fit or to be constrained to specified values or
within a certain range.
The code returns the best fit values for the stellar kinematics and
the best fit spectra of the two stellar components, which will be used
in Section \ref{sec:ssp} to derive their stellar populations.  Also,
the code returns the light contribution of one component relative to
the total galaxy spectrum. If we label the two components with ``A''
and ``B'', the code returns $F_A = \frac{I_A}{I_A+I_B}$, where $I_A$
and $I_B$ are the flux of the two components. By construction, $F_B =
1-F_A$.

Our technique has been successfully applied to disentangle
counter-rotating stellar disks in galaxies \citep{Coccato+11,
  Coccato+13, Pizzella+14}, stars from the bulge and disk
\citep{Fabricius+14}, and stars from the host galaxy and the
surrounding polar disk \citep{Coccato+14}.

For the special case of NGC 4191, in order to obtain a reasonable and
stable fit, it is necessary to constrain the flux ratio of the two
kinematic components by performing an independent photometric
decomposition (Section \ref{sec:photometry}). This is due to a
degeneracy between the parameters that describe the kinematics and the
stellar populations of the two components. The same approach has been
adopted in \citet{Katkov+11} and \citet{Coccato+14}.

\subsubsection{Photometric contraints}
\label{sec:photometry}

The aim of this section is to perform a photometric decomposition of
NGC 4191 in order to constrain the flux parameter $F_A$ in the spectral
decomposition code.

We acquired a $g$-band image (the closest match to the VIRUS-W
wavelength range) from the SDSS DR7 archive and analyzed it using the
2D image-fitting code \textsc{Imfit} \citep{Erwin+15}. The fitting
procedure includes the convolution for a Moffat point spread function
(PSF) computed using the median FWHM and $\beta$ values measured from
stars in the same image.

The best fit model is obtained adopting three components: a bulge with
S\'ersic profile, one inner exponential disk, and one outer
exponential disk.  In the fitting procedure, the center has been kept
fixed for all the 3 components. The position angle and the ellipticity
of the three components are independent, but constant with radius.

\begin{figure}
\begin{center}
\psfig{file=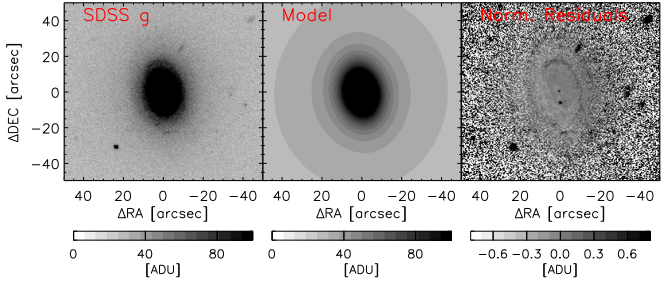,width=9cm,clip=}
\caption{Result of the photometric decomposition of NGC 4191.  Left
  panel: SDSS $g$ image; central panel: best fit model; right panel:
  normalized residuals (data - model)/model. North is up and East is
  left.}
\label{fig:photom1}
\end{center}
\end{figure}

\begin{figure}
\begin{center}
\psfig{file=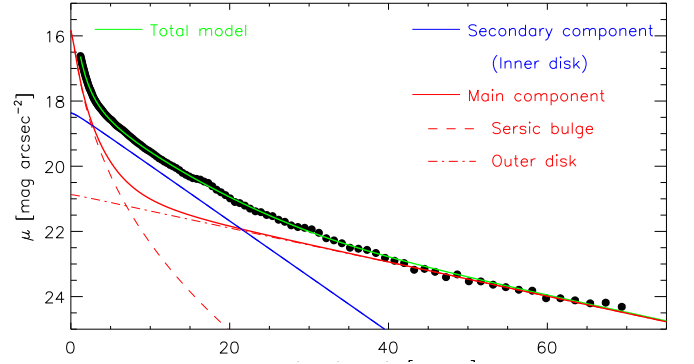,width=9cm,clip=}
\caption{Results of the photometric decomposition. Filled circles:
  surface brightness radial profile measured on the SDSS $g$-band
  image with the iraf task ellipse. The blue curve is the best fitting
  inner disk component, the red-dashed curve is the best fitting
  S\'ersic bulge, and the red dot-dashed curve is the outer disk
  component. The green curve is the sum of all the best fitting
  components. Red and blue curves represent the profiles of the
  adopted definition for the main and secondary kinematics components,
  respectively.}
\label{fig:photom2a}
\end{center}
\end{figure}

The structural parameters of the best fitting components are given in
Table \ref{tab:photom}, and the best fit model is shown in Figures
\ref{fig:photom1} and \ref{fig:photom2a}. The residual map of
Fig. \ref{fig:photom1} highlights a spiral-arm pattern, indicating the
presence of faint spiral arms and dust in this galaxy, despite of its
classification as S0.

Our photometric decomposition represents an improvement of the
previous single component model by \citealt{Krajnovic+13}. One way to
evaluate the \textit{relative} goodness-of-fit of different models to
the same data is via information-theoretic measures such as the Akaike
Information Criterion (AIC) or Bayesian Information Criterion (BIC),
both of which are output by \textsc{Imfit} at the end of the fitting
process. Unlike the $\chi^{2}$ statistic, AIC and BIC do not require
that models being compared be nested. The absolute values of these
statistics are irrelevant; what matters is the relative value, with
lower values indicating a better match to the data. Traditionally,
$\Delta{\rm AIC} = {\rm AIC}_{1} - {\rm AIC}_{2} < -10$ is taken as an indication
for the clear superiority of model 1 over model 2. We find that the
S\'ersic + two exponential model is clearly favored. In the case of
the $g$-band image, the three-component model has $\Delta{\rm AIC} =
-1442$ compared to the S\'ersic + single-exponential model and
$\Delta{\rm AIC} = -15006$ compared to the single-S\'ersic model. The
values for the BIC are similarly large: $\Delta{\rm BIC} = -1398$ and
$-14919$, respectively.

\begin{figure}
\begin{center}
\psfig{file=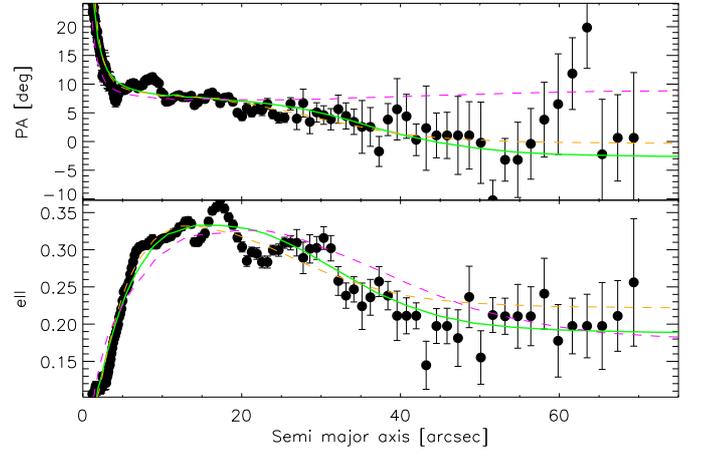,width=9cm,clip=}
\caption{Comparison between the radial profile of ellipticity (top
  panel) and position angle (bottom panel) obtained with different
  photometric models. Filled circles: surface brightness radial
  profile measured on the SDSS $g$-band image with the iraf task
  ellipse. Green line: three-components model; orange dashed line:
  two-components model; magenta dashed line: single-component model.}
\label{fig:photom2b}
\end{center}
\end{figure}

Figure \ref{fig:photom2b} provides additional support to the choice of
three-components model over one- or two- components models. It shows
the radial profiles of ellipticity and position angle as measured from
the $g$-band image and from the various models.  We can see that the
single-S\'ersic model does a particularly poor job of reproducing the
position angle radial variation, and the two-components model does not
reproduce the change of ellipticity as nicely as the three-components
model for $25''<a<30''$ and $a>45''$.  For use in constraining the flux
parameter $F$ in the spectral decomposition, we therefore use the
three-components model.

\begin{table}
\begin{center}
\caption{Best fit parameters of the photrometric decomposition of NGC 4191.}
\begin{tabular}{l r c l l}
\hline
\hline
\multicolumn{5}{c}{S\'ersic bulge}\\
$PA$	&	 13.3  &$\pm$&  0.7 & [degrees]\\
$ell$   &	 0.128 &$\pm$& 0.003& \\
$n$	&	2.87  &$\pm$& 0.06& \\
$\mu_e$	&	19.7  &$\pm$& 0.05 &  [mag arcsec$^{-2}$]\\
$r_e$	&	2.22  &$\pm$& 0.07&  [arcsec]\\
$m_{\rm TOT}$&	14.89 &$\pm$& 0.05  &  [mag] \\
\hline
\multicolumn{5}{c}{Inner exponential disk}\\
$PA$	&	7.8    &$\pm$& 0.1&  [degrees] \\
$ell$	&	0.384 &$\pm$& 0.003 & \\
$\mu_0$	&	19.78  &$\pm$& 0.02 & [mag arcsec$^{-2}$]\\
$h$	&	6.54 &$\pm$& 0.05 & [arcsec]\\
$m_{\rm TOT}$&	14.23 &$\pm$& 0.02  &  [mag] \\
\hline
\multicolumn{5}{c}{Outer exponential disk}\\
$PA$	&	$-$2.7  &$\pm$& 0.7 &  [degrees] \\
$ell$	&	0.189 &$\pm$& 0.004 & \\
$\mu_0$	&	22.26 &$\pm$& 0.01  & [mag arcsec$^{-2}$] \\
$h$	&	20.9  &$\pm$& 0.1  &  [arcsec] \\
$m_{\rm TOT}$&	14.02 &$\pm$& 0.01  &  [mag] \\
\hline
\hline
\end{tabular}
\label{tab:photom}
\end{center}
\end{table}

We define ``main kinematic component'' the sum of the bulge and outer
disk and ``secondary kinematic component'' the inner disk.  The main
kinematic components contains the $\sim 64\%$ of the total galaxy
luminosity, whereas the secondary kinematic component contains the
$\sim 36\%$ of the total galaxy luminosity.
The particular association of the inner disk component with the
counter rotating kinematic component is driven by its relative local
luminosity. From the observed peaks in the LOSVDs we know
that the two components have flux ratios between 1:3 and 1:1. Other
combinations, such as defining the secondary component as the sum of
the two disks, would have set one of the components below a detection
limit ($F<0.15$, see Appendix \ref{sec:errors}), which is inconsistent with the observations.

At each position $(x,y)$ in the field of view, we can define the
fraction of galaxy surface brightness contributed by the main and
secondary components as:

\begin{eqnarray}
F_{\rm main}(x,y) &=& F_A = \frac{I_{\rm main}(x,y) }{I_{\rm main}(x,y) + I_{\rm second}(x,y)} \nonumber \\
                &=&  \frac{I_{\rm bulge}(x,y) + I_{\rm outer\ disk}(x,y)}{I_{\rm bulge}(x,y) + I_{\rm outer\ disk}(x,y)+ I_{\rm inner\ disk}(x,y)} \nonumber \\
F_{\rm second}(x,y) &=&  F_B = \frac{I_{\rm second}(x,y) }{I_{\rm main}(x,y) + I_{\rm second}(x,y)}  \\
                   &=&  \frac{I_{\rm inner\ disk}(x,y)} {I_{\rm bulge}(x,y) + I_{\rm outer\ disk}(x,y)+ I_{\rm inner\ disk}(x,y)} \nonumber \\
                  &=&  1-F_{\rm main}(x,y)
\end{eqnarray}

\begin{figure}
\begin{center}
\psfig{file=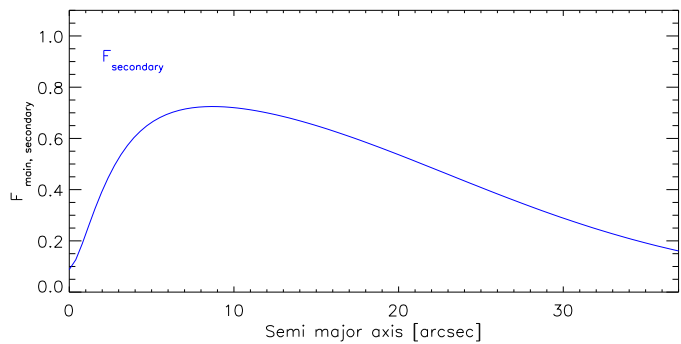,width=9cm,clip=}
\caption{Radial profiles of the relative flux contribution of the
  secondary stellar components $F_{\rm secondary}$. The contribution
  of the main component is $F_{\rm main} = 1-F_{\rm secondary}$
  by construction.}
\label{fig:photom3}
\end{center}
\end{figure}

In Figure \ref{fig:photom3} we show the radial trend along the major
axis of $F_{\rm second}$. 

The results of the photometric decomposition can be now used to
constrain the spectral decomposition analysis. For a given spatial
bin, we constrain the parameter $F_{\rm A}$ in the spectral
decomposition code to be within the range $F_{\rm main} -0.05 < F_{\rm A} < F_{\rm main}
+ 0.05$; $F_{\rm main}$ is given by:
\[
F_{\rm main} = \frac{\sum I_{\rm main}(x,y)}{\sum I_{\rm main}(x,y)
  + \sum I_{\rm second}(x,y)} 
\] 

where the sum is done over all the spaxels contributing to a given
spatial bin. 

\subsubsection{Kinematics of the counter-rotating stellar components}

\begin{figure*}
\begin{center}
\psfig{file=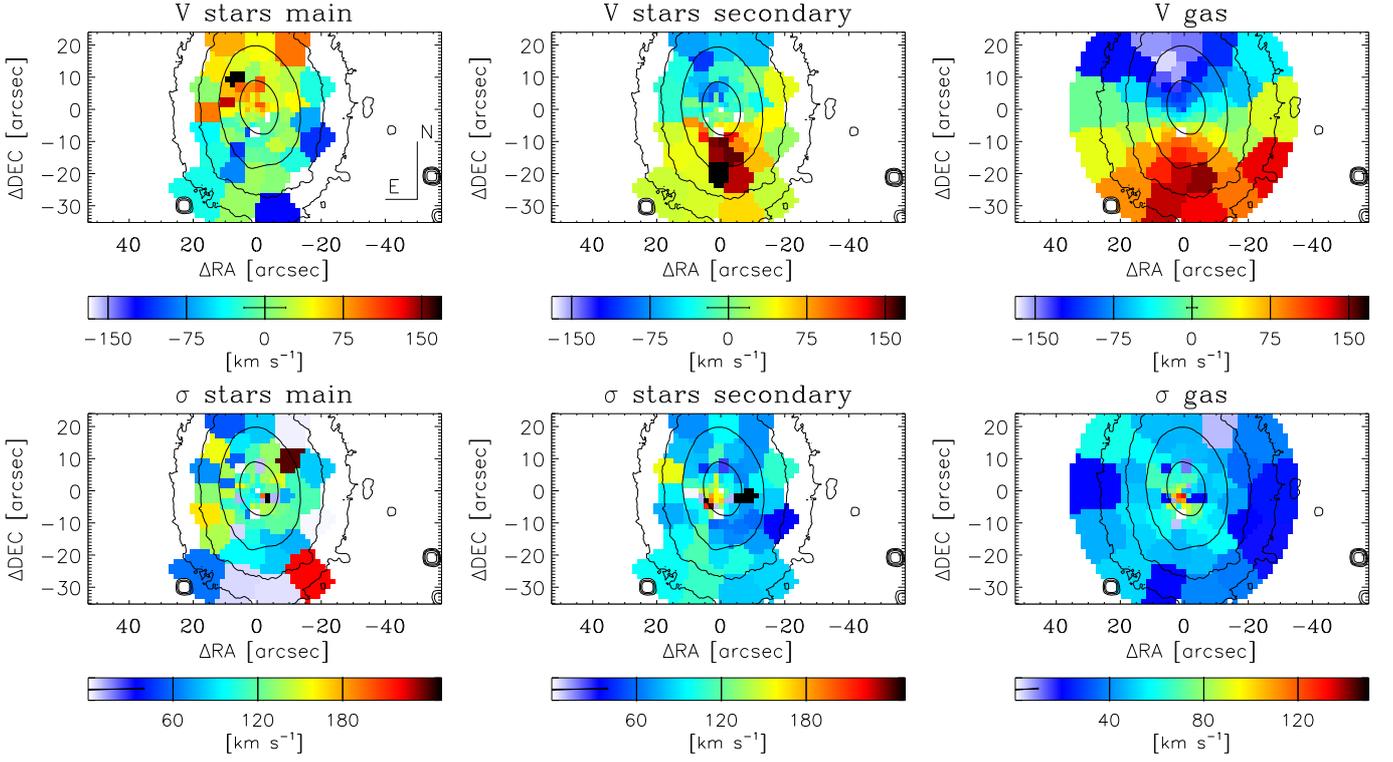,width=18.cm,clip=,bb=20 350 730  750}
\caption{Result of the spectral decomposition of NGC 4191. Velocity
  (top panels) and velocity dispersion (bottom panels) maps for the
  main stellar component (left panels), secondary counter-rotating
  stellar component (central panels) and counter-rotating ionized-gas
  component (right panels) are shown. Spatial bins with $S/N$ lower
  than 25 per \AA\ are not shown for the stellar components.  The
  horizontal bars in the color scale at the bottom of each panel
  indicate the mean error.}
\label{fig:two_comp}
\end{center}
\end{figure*}

Figure \ref{fig:two_comp} shows the kinematics of the two
stellar components, as obtained from the spectral decomposition fit.

The main galaxy component, which consists in the combination of the
S\'ersic bulge and the large-scale exponential disk, is more
luminous. It is characterized by a rotation amplitude of $\sim 70$
\kms, with some scatter, and an average velocity dispersion of $\sim
120$ \kms.

The secondary galaxy component, which is represented by the inner
exponential disk, is less luminous and it counter-rotates with respect
the main stellar component. It has  a rotation amplitude of
$\sim 130$ \kms\ and an average velocity dispersion of $\sim 90$
\kms. The direction of rotation is aligned with that of the ionized gas.

As additional test, we have performed the spectral decomposition using
the constrains from the two-components photometric model. Although the
main kinematic results are consistent with those presented above, the
two-dimensional kinematic maps obtained with the two-components
photometric model have higher noise. This is due to the poorer quality
of the photometric fit, which translates into larger uncertainty in
constraining the light content of the counter-rotating stellar
components.

\subsection{Stellar populations of the counter-rotating stellar
  components}
\label{sec:ssp}

We measured the \hb, \mgb, \fei, and \feii\ line-strength indices on
the optimal template of each stellar component returned by the
spectral decomposition code. The spectra are broadened with a Gaussian
function to match the spectral resolution of the Lick system (8.4 \AA,
\citealt{Worthey+97}) . Therefore, we obtained two sets of indices in
each spatial bin, one set for each kinematic component.

Measurements were scaled to the Lick system by comparing the line
strength indices measured in the spectra Lick standard stars HD102870
and HD125560 obtained with VIRUS-W with those published by
\citet{Worthey+94}. The offsets are $-0.35 \pm 0.13$ \AA\ (\hb),
$-0.05 \pm 0.01$ \AA\ (\mgb), $-0.02 \pm 0.15$ \AA\ (\fei), and $-0.06
\pm 0.12$ \AA\ (\fei). We applied this correction to the \hb\ and
\mgb\ indices, but not for the metal indices, because they are
negligible considering the measurements errors.

\begin{figure}
\begin{center}
\psfig{file=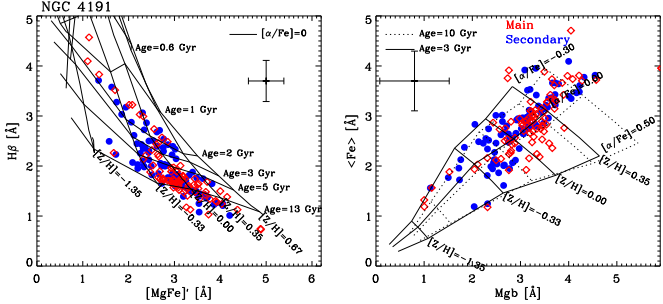,width=9cm,clip=}
\caption{Line strength indices of the two counter-rotating stellar
  components in NGC 4191. Blue filled circles and red open diamonds
  refer to the main and secondary component, respectively. The error
  bars represent the mean errors in the measurements. The line grids
  represent the predictions of simple stellar population models
  \citep{Thomas+03}. Spatial bins with $S/N$ lower than 25 per \AA\
  are not considered.}
\label{fig:indices}
\end{center}
\end{figure}

In Figure \ref{fig:indices}, we show the combined indices \hb, \mgb\,
\mgfe, and \mfe\ of the two stellar components.
The scatter in the plots is large, reflecting the uncertainties in our
measurements. In particular, we note that the scatter in the
measurements in the \mgb\, \mfe\ plane of Fig. \ref{fig:indices} is
larger for the secondary component. This is probably due to a combined
effect of i) radial gradients that are more pronounced for the
secondary component (see below); and ii) secondary component being on
average the less luminous, and therefore the errors in the
measurements are larger.

\begin{figure*}
\begin{center}
\psfig{file=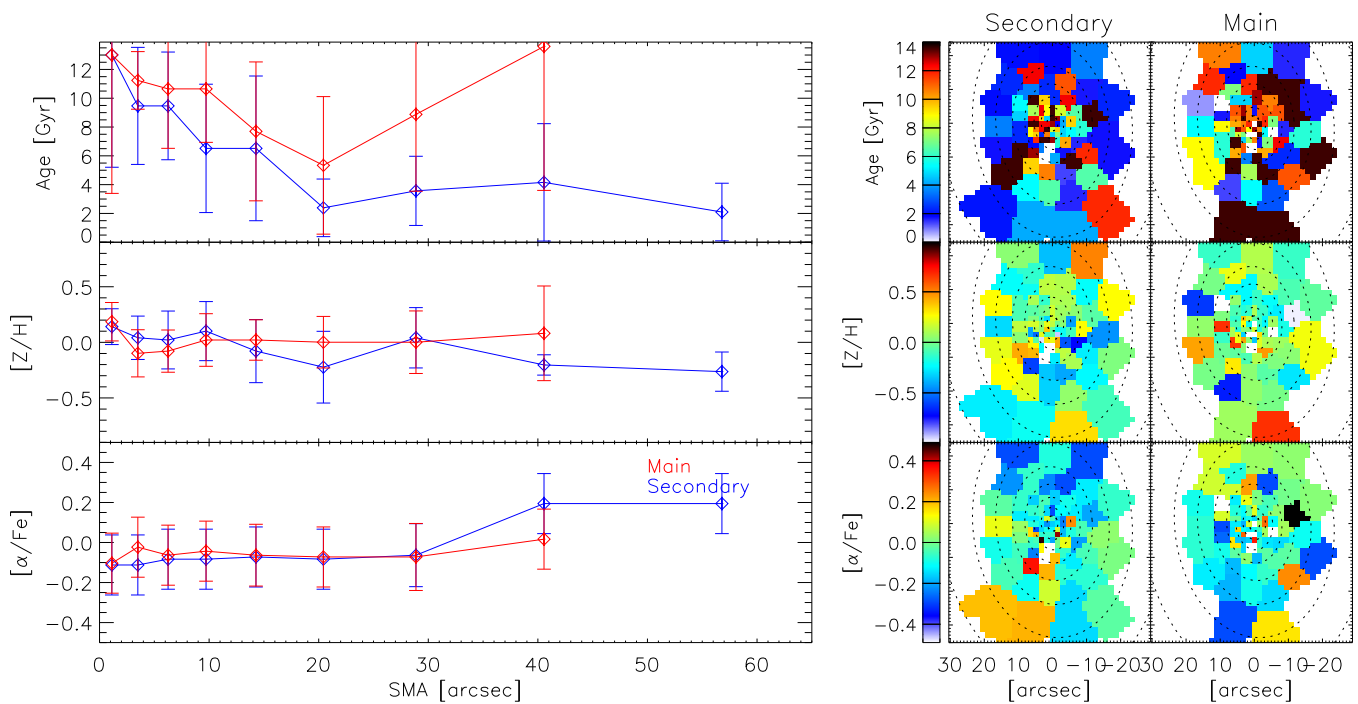,width=18cm,clip=}
\caption{Properties of the stellar populations of the counter-rotating
  components in NGC 4191. Left panels: median values of age (top
  panel), [Z/H] (middle panel) and [$\alpha$/Fe] (bottom panel) within
  elliptical anuli for the main (red line) and secondary (blue line)
  stellar components. Right panels: two dimensional maps of the
  stellar population properties. The dashed lines indicate the
  elliptical anuli used in the computation of the median values
  presented in the left panels. Spatial bins with $S/N$ lower than 25
  per \AA\ are not considered in the analysis.}
\label{fig:ssp}
\end{center}
\end{figure*}

Fig. \ref{fig:indices} also compares our measurements to the
predictions of stellar population models \citep{Thomas+03}. This
allows us to derive their two-dimensional maps of age, metallicity,
and abundance ratios of $\alpha$ elements, which are shown in Figure
\ref{fig:ssp}. Figs. \ref{fig:indices} and \ref{fig:ssp} do not show
the results on those spatial bins where the $S/N$ is lower than 25 per
\AA.
In Fig. \ref{fig:ssp} we also compute median values within elliptical
anuli to highlight the presence of radial gradients in the properties
of the stellar populations. The error in each elliptical bin is
computed as the maximum between i) the scatter of the measurements
within that elliptical bin, and ii) the weighted mean error
$1/\sqrt{\sum_i 1/\sigma_i^2}$, where $\sigma_i$ are the individual
measurements errors in age, [Z/H], and [$\alpha$/Fe].

The properties of the stellar populations of the two counter-rotating
stellar components are very similar, as  shown by
  Figs. \ref{fig:indices} and \ref{fig:ssp}. Because the scatter in
  the two-dimensional maps is large and a direct comparison is
  difficult, we compare average quantities. Metallicities and
abundances of $\alpha$ elements are consistent with solar, without
significant radial gradients. Their mean values are:
$\langle$[Z/H]$\rangle_{\rm main} = 0.02 \pm 0.09$ dex,
$\langle$[Z/H]$\rangle_{\rm second} = -0.1 \pm 0.2$ dex,
$\langle$[$\alpha$/Fe]$\rangle_{\rm main} = -0.05 \pm 0.05$ dex, and
$\langle$[$\alpha$/Fe]$\rangle_{\rm second} = -0.02 \pm 0.05$ dex.
On the other hand, the age radial profiles show a mild indication of radial
  gradients.
The age of the stars in the main component ranges from $\sim 12$ Gyr
in the central regions down to $\sim 6$ Gyr at $21''$. Outside
$21''$ the age rises again, but the uncertainties at these radii
become too large to derive conclusive results. One possible
explanation is that, if the observed raise in age is real, it might
be due to accretion of old stars, which were accreted in the same
rotation direction as the main component.
The age of the stars in the secondary component ranges from $\sim 12$
Gyr in the central regions down to $\sim 3$ Gyr at $21''$, remaining
nearly constant afterwards. We compare the mean ages of the two
components within $21''$, which can be considered the upper limit for
trusting age measurements in the main component. We find:
$\langle$age$\rangle_{\rm main}(R<21'') = 10 \pm 3$ Gyr and
$\langle$age$\rangle_{\rm second}(R<21'') = 5 \pm 4$ Gyr.
The mean ages of the two stellar components agree within the errorbars; on
the other hand, the radial profiles in Fig. \ref{fig:ssp} seem to
indicate that the secondary component is systematically younger than
the main galaxy by $\approx 2$ Gyr at all radial bins. Implications of
this age difference, if real, will be discussed in Section
\ref{sec:summary}.

By comparing the mean properties of the stellar populations with the
model predictions by \citet{Maraston+05} assuming a Salpeter initial
mass function, we can derive the mean values of the stellar
mass-to-light ratios of the two components in the $g$ band: $\langle
M/L_{\ast} \rangle_{\rm main} = 6.7 \pm 2.5$,  $\langle M/L_{\ast}
\rangle_{\rm second} = 3.3 \pm 2.5$. These informations, combined with
the luminosity of the two components measured in
Sect. \ref{sec:photometry} and the adopted distance of 42.4 Mpc, allow
us to estimate their masses: $M_{\rm main} = 6.2 \pm 2.3 \cdot 10^{10}$ 
\Msun\ and  $M_{\rm second} = 1.0 \pm 9.4 \cdot 10^{10}$\Msun; the main
stellar component is  $\approx 6$ times more massive than the
secondary component.

\subsection{Kinematics and distribution of the ionized gas}
\label{sec:ionized_gas}

The spectral decomposition routine fits also the ionized gas emission
lines. The gas component (see Fig. \ref{fig:two_comp}) rotates along
the same direction of the secondary stellar component with an
amplitude of $\sim 150$ \kms. The velocity dispersion decreases with
radius: from $\sim 60$ \kms\ in the center down to $\sim 30$ \kms\ in
the outskirts. The high value of velocity dispersion measured in the
inner 1'' ($\sim 100$ \kms) is probably due to the seeing and the
limited spatial resolution.

The emission lines are very weak, therefore it was necessary to
spatially bin the data to measure their kinematics as in the case of
the stars. For convenience, we adopted the same spatial binning as for
the stellar component to measure the gas kinematics. However, this
``coarse'' binning, prevents us to investigate the spatial
distribution of the emission lines. We therefore re-analyzed the gas
distribution on a finer spatial grid and measured the equivalent widths
of the \hb\, \oiiip, and \oiiig\ emission lines on this latter spatial
binning.

\begin{figure}
\begin{center}
\psfig{file=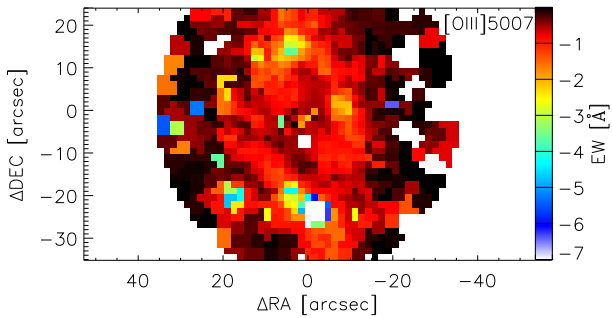,width=8cm,clip=}
\caption{Distribution of the \oiiig emission line. The gas is
  distributed on a ring-like structure aligned with the photometric
  major axis of the secondary component (see text for details).}
\label{fig:gas_distribution}
\end{center}
\end{figure}

Figure \ref{fig:gas_distribution} shows the spatial distribution of
the \oiiig\ emission line equivalent widths, which is the most
intense. Although the outer regions are dominated by the noise due to
lower signal-to-noise of the data, an elongated ring-like structure
(or maybe a spiral arm structure) with three blobs is visible
within $25''$. The position angle of the structure, using the two
brightest blobs in the equivalent width map as reference, is
8.0$^{\circ}$, which is very close to the position angle of the
secondary component, i.e. 7.8$^{\circ}$ (Table \ref{tab:photom}). This
strengthens the association of the ionized-gas component to the
secondary stellar component. 

Interestingly, a similar spiral-like structure with same size and
orientation, is observed in the residual map of
Fig. \ref{fig:photom1}, consistently with the concentration of gas
along spiral arms. Unfortunately, the spatial binning we adopted for
the stellar population studies does not have enough spatial resolution
to resolve the spiral arms and separate their stellar population from
that of the rest of the galaxy.

It is interesting to compare, although qualitatively, the distribution
of emission lines of NGC 4191 with those of other stellar
counter-rotating galaxies studied so far. The distribution of the
ionized gas in NGC 4191 resembles those of NGC 3593 and NGC 5719, where
the gas is also distributed along a ring-like structure with presence
of blobs. In addition, in the case of NGC 5719, the location of the
most intense blobs coincides with that of the youngest stars.  In the
case of NGC 4550, the gas has irregular distribution, with no
signature of ring-like or spiral-like structures. Also, NGC 4550 is
the galaxy where the two stellar components are the oldest.

\section{Discussion and conclusions}
\label{sec:summary}

The LOSVD of NGC 4191 are complex and suggest the presence of two
kinematically-decoupled stellar components. Their properties have been
investigated with the aid of a non-parametric fit \citep{Fabricius+14}
and a spectral decomposition technique \citep{Coccato+11}.

We also performed a parametric photometric decomposition of the
surface-brightness distribution of NGC 4191. Our detailed analysis
reveals the presence of three structural components, which were
overlooked in previous studies: a S\'ersic bulge and two exponential
disks of different scale lengths. Also, faint spiral structures (stars
and dust) are visible from the residual map.

We then linked the photometric components to the kinematic components
unveiled by the spectroscopic analysis. The intensity of the double
peaks in the LOSVDs allowed us to associate the bulge and the outer
exponential disk with the main kinematic component. Together they
represent the $\sim 64\%$ of the total galaxy luminosity.
Consequentially, we associated the inner exponential disk with the
secondary kinematic component, which represents the $\sim 36\%$of the
total galaxy light.

The information on the photometry allowed us to constrain the relative
light contribution of the two kinematic components in the spectral
decomposition fit. This decreases the degeneracy between the various
fitting parameters that characterize the two components (kinematics,
surface brightness, and stellar populations) and allows us to obtain a
more reliable fit.

We find evidences that the two kinematic components in NGC 4191
counter-rotate with respect to each other. The secondary component
rotates faster and along the same direction as the ionized gas. The
main component rotates slower and it has higher velocity dispersion,
consistent with hosting also bulge stars.

The properties of the stellar populations of the two kinematic
components are very similar, although the large uncertainties in
  the measurements prevent us to make strong claims. We have
  azimuthally averaged the measurements onto elliptical bins to reduce
  the statistical noise and highlight the presence of radial
  gradients. Fig. \ref{fig:ssp} gives a tentative indication that
  the luminosity-weighted ages of the stars in the secondary disk are
systematically lower than those of the main components at all
radii. We can explain the age difference between the two
  components with the following scenario.

The combination of kinematics and stellar populations gives us clues
on the formation mechanism of NGC 4191. Our data favor the scenario
in which a disk galaxy (represented by the main kinematic component,
i.e., the bulge plus the outer disk) acquired material (either
gas/stars accretion or galaxy mergers) from the outside onto
retrograde orbit, which formed the counter-rotating secondary
component. At first glance, this scenario is the same of those of the
other counter-rotating systems studied so far \citep{Coccato+11,
  Coccato+13, Katkov+13, Pizzella+14}. However, there are some
important differences between NGC 4191 and these other galaxies.

First, the age radial gradients observed in NGC 4191 are much steeper
than those observed in other systems. Moreover, the age gradient of
the secondary component is negative, whereas in the other systems is
positive. This is indicative of an inside-out star formation process
for the secondary disk, and favors the gas accretion scenario over
 major merger. A nearly 1:1 merger in fact would have redistributed the
stars along all directions, smoothing any significant age gradient.

Second, there are almost no differences in the metallicity and
$\alpha$-elements overabundance ratio between main and secondary components,
whereas differences were observed in other galaxies. In other
galaxies, the scenario that better explained the differences in [Z/H]
and [$\alpha$/Fe] between the main and secondary components is gas
accretion from cosmic gas reservoir or from a nearby gas-rich
companion. The case of NGC 5719 offers the best example of such a
scenario: we observe the on-going gas stripping process from its
nearby companion that forms the counter-rotating stellar disk
\citep{Vergani+07, Coccato+11}. On the contrary, the data for NGC 4191
suggest that the accreted gas has origin in common with the gas that
formed the main galaxy component.

The observed properties of NGC 4191 can be interpreted in light of
cosmological simulations \citep{Algorry+14}. The main galaxy component
is formed via accretion of gas from two filaments  that have very
  similar chemical composition. As time passes, each filament torques
the other in opposite direction, resulting into a ``double accretion''
with opposite spin. Initially, the accretion from the two filaments
occur at the same time. The collisional nature of the gas ensures that
only one stellar component forms, from gas whose properties is a
mixture of that of the two filaments, and with the spin dictated by
the most massive filament. After a while ($\sim 2$ Gyr, considering
the measured age difference between the two components in each radial
bin), the accretion along one of the filaments stops.

The accretion process continues along the counter-rotating filament,
which wipes out the remaining gas and forms the secondary
counter-rotating component. The star formation occurs inside-out, as
suggested from the negative age gradient; the new born stars have
properties very similar to those of the main component, because the
gas is a mixture of that of the second filament and what remained from
the first filament. Moreover, the formation of the secondary component
must have occurred right after the formation of the main component and
very rapidly, without leaving the stars of the first component enough
time to reprocess and enrich the gas with metals and/or $\alpha$
elements. This is consistent with the relative small age difference
between the two components and their equal and constant [Z/H] and
[$\alpha$/Fe] radial profiles. The rapid star formation process and
the solar abundances and metallicity measured for the stellar
populations, imply that the infalling gas was already enriched,
because otherwise there would not be enough time to reach the measured
values of [Z/H] and [$\alpha$/Fe].

The hypothesis where the main component formed first is consistent
with the observed properties of the photometric components: the
position angles are slightly different and the ellipticities of the
S\'ersic bulge and outer disk are smaller than that of the secondary
component (inner disk, see Tab. \ref{tab:photom}). The most likely
interpretation is that the bulge is triaxial and the disks have
different thickness (the external disk being the thickest). This is in
agreement with the proposed formation scenario: the main component
formed first and had more time to heat up its disk during the
accretion of the gas that originated the secondary component. Gas
accretion can indeed heat the host stellar disk during the formation
of a counter-rotating component (i.e., \citealt{Thakar+96}).

To conclude, we would like to remark the importance of disentangling
and measuring the properties of individual components in multi-spin
galaxies to understand their formation mechanisms. Also, the special
case of NGC 4191 supports the validity of the double peaked signature
in the velocity dispersion field associated to zero-velocity rotation
as selection criteria for identifying counter-rotating galaxies. This
is very useful for large integral field spectroscopic surveys to
provide a statistically completed census of counter-rotating galaxies
in the nearby universe.


\bibliography{coccato2015}

\begin{appendix} %
\section{Errors of the spectroscopic decomposition}
\label{sec:errors}

The ability of the spectral decomposition code to measure the
kinematic and properties of the stellar populations depends on many
parameters. First, on the differences of the two stellar components:
the more separated they are in kinematics and stellar populations, the
easier is to decouple their contribution from the observed galaxy
spectrum. Also, the spectral resolution and wavelength range of the
instrument and the signal-to-noise of the observations are important.

We therefore simulate galaxy spectra that represent observations and
the instrumental set-up as close as possible. We then perform the
spectral decomposition on those simulated spectra to derive the errors
on the measured quantities.

The simulated galaxy spectra are constructed by creating and adding
two stellar templates broadened to have 100 \kms\ of velocity
dispersion. The stellar templates are both constructed using the same
star, HD106210, observed with the VIRUS-W (to match the instrumental
set-up), that has line strength indices as close as possible to the
mean measured values.

We then explore the parameters space defined by the velocity
separation of the two component ($\Delta V$), and the signal-to-noise
ratio ($S/N$). For each point $(\Delta V, S/N)$ in the parameter space
we construct a set of 100 simulated galaxy spectra by adding Gaussian
noise. We then perform the spectral decomposition on those simulated
spectra to derive the errors on the measured quantities.  The
parameter space is sampled by $\Delta V= 50,75,100,150,250$ \kms, and
$noise=0.005, 0.02, 0.03, 0.10$, that correspond to $S/N$ ranging
from 480 to 25 per \AA\ (the mean $S/N$ of the binned spectra is 90
per \AA) \footnote{We define the noise (rms) as the standard
  deviation of the residuals between the galaxy observed spectrum
  (normalized to its median value) and the best fit model. The actual
  computation on the wavelength region used in the fit and with the
  robust\_sigma IDL routine. This value of rms is converted into a
  proxy for the $S/N$ per angstrom by $S/N=1/{\rm rms}/\sqrt{0.178}$,
  where 0.178 is the inverse dispersion in \AA\ pixel$^{-1}$.}. 

We find that the errors on the absorption line indices does not vary
much with velocity separation, because the stellar populations of the
two components used in the simulations are the same (i.e. there is no
degeneracy with the kinematics). There is also little dependence with
$S/N$. The mean errors on the indices are 0.2 \AA\, 0.4 \AA, 0.2 \AA,
and 0.2 \AA, for the \hb, \mgb, \fei, and \feii\ indices,
respectively.

\begin{figure}
\begin{center}
\vbox{
   \psfig{file=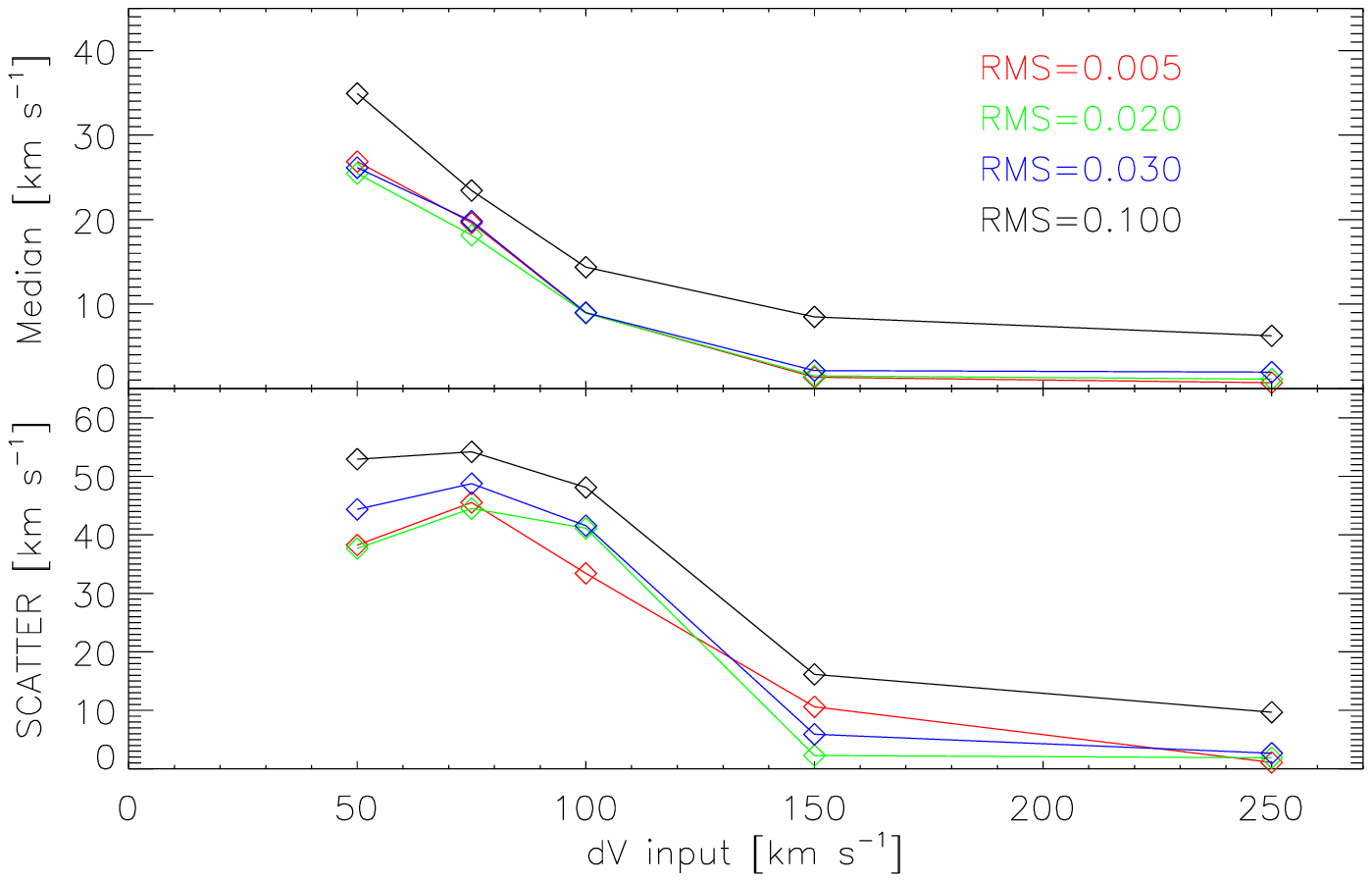,width=9cm,bb=55 392 465 658, clip=}
   \psfig{file=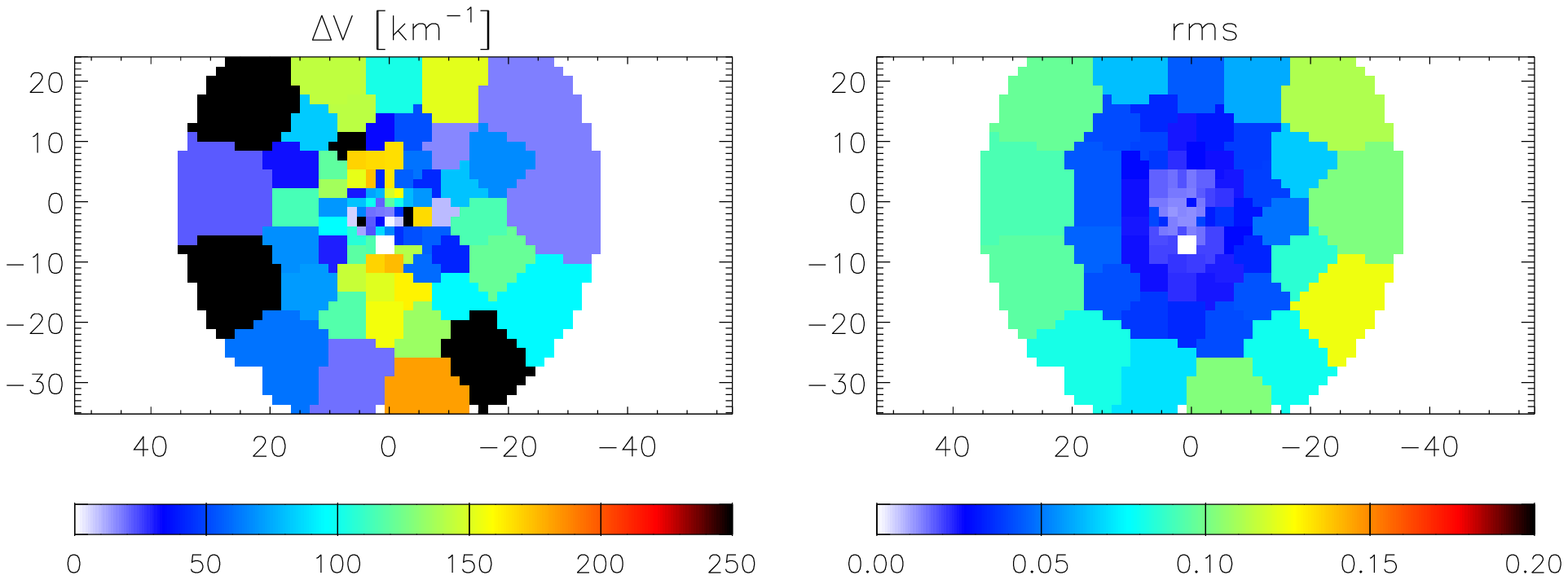,width=4cm,angle=90, clip=,bb= 28 226 294 792}
}
\caption{Top figure: median value (top panel) and standard deviation
  (bottom panel) of the error on the recovered velocities as function
  of the velocity difference between the two simulated stellar
  components. Different lines represent different signal-to-noise
  values per \AA: 480 (rms = 0.005), 120, 80, and 24 (rms =
  0.1). Bottom figure: two-dimensional maps of the absolute velocity
  difference between the two stellar components (left panel) and the
  fit rms (right panel).}
\label{fig:sim_vel} 
\end{center}
\end{figure}

We find that the errors on velocity separation between the two
components depend on the input velocity separation and on the
$S/N$. Figure \ref{fig:sim_vel} shows the error on the recovered
velocity separation as function of the input noise and input velocity
separation. For typical values measured along the galaxy kinematic
major axis, where the velocity separation is $\Delta V>150$\kms\ and
the noise is ${\rm rms}<0.05$, the error on the velocity separation
between the two components is $\simeq 30$\kms.  Also, the error on the
recovered flux of the two components is $\sim 15$\%.

\end{appendix}

\end{document}